\documentclass{article}
\usepackage[utf8]{inputenc}

\usepackage{float}
\usepackage{multirow}
\usepackage{graphicx}
\usepackage[font=footnotesize]{subcaption}
\usepackage[labelfont=bf]{caption}
\usepackage{bm}
\usepackage{booktabs}
\usepackage[svgnames,table]{xcolor}
\usepackage{pifont}

\usepackage{amssymb}
\usepackage{amsbsy}
\usepackage{amsmath}
\usepackage{listings} 
\usepackage{mathrsfs}

\usepackage[backend=biber,style=authoryear,maxbibnames=99,natbib,uniquename=false]{biblatex}
\usepackage{appendix}
\usepackage[affil-it]{authblk}

\usepackage[a4paper, margin=1in]{geometry}
\usepackage{setspace}
\usepackage{blindtext}
\usepackage{boldline}
\makeatletter
\def\Cline#1#2{\@Cline#1#2\@nil}
\def\@Cline#1-#2#3\@nil{%
  \omit
  \@multicnt#1%
  \advance\@multispan\m@ne
  \ifnum\@multicnt=\@ne\@firstofone{&\omit}\fi
  \@multicnt#2%
  \advance\@multicnt-#1%
  \advance\@multispan\@ne
  \leaders\hrule\@height#3\hfill
  \cr}
\makeatother

\usepackage{fancyhdr}
\title{ \vspace{-2.69cm} \Large 
Assessing Look-Ahead Bias in Stock Return Predictions Generated By GPT Sentiment Analysis}
\author[1]{Paul Glasserman}
\author[2]{Caden Lin}
\affil[1]{%
  \small Columbia Business School, Decision, Risk, and Operations Division}
\affil[2]{%
  \small Columbia University, Department of Mathematics}
\date{September 2023}

\begin{document}
\maketitle

\begin{abstract}
    \normalsize
Large language models (LLMs), including ChatGPT, can extract profitable trading signals from the sentiment in news text. However, backtesting such strategies poses a challenge because LLMs are trained on many years of data, and backtesting produces biased results if the training and backtesting periods overlap. This bias can take two forms: a \emph{look-ahead bias}, in which the LLM may have specific knowledge of the stock returns that followed a news article, and a \textit{distraction effect}, in which general knowledge of the companies named interferes with the measurement of a text's sentiment. We investigate these sources of bias through trading strategies driven by the sentiment of financial news headlines. We compare trading performance based on the original headlines with de-biased strategies in which we remove the relevant company's identifiers from the text. In-sample (within the LLM training window), we find, surprisingly, that the anonymized headlines outperform, indicating that the distraction effect has a greater impact than look-ahead bias. This tendency is particularly strong for larger companies --- companies about which we expect an LLM to have greater general knowledge. Out-of-sample, look-ahead bias is not a concern but distraction remains possible. Our proposed anonymization procedure is therefore potentially useful in out-of-sample implementation, as well as for de-biased backtesting.\end{abstract}
\pagebreak

\section{Introduction}

\subsection{Background}


Many studies have found that news stories can be used to predict short-term market returns through sentiment analysis and other natural language processing tools; see, e.g.,
\citet{das2007, tetlock2007, tetlock2008, calomiris2019, ke2019, garcia2023, glasserman2022}.
Work in this area has mostly relied on fixed lexicons for sentiment analysis or used predictive techniques that can be retrained on a rolling window of data.


More recently, large language models (LLMs) like the Generative Pre-trained Transformer (GPT) architecture are proving to be valuable tools for processing news text to forecast market responses, despite not being trained specifically for this task; see, in particular, \citet{lopezlira2023chatgpt} and also \citet{Chen_2023} and \citet{hansen2023}. LLMs are trained on massive datasets spanning many years. This feature, which is central to their success, poses a challenge for designing and backtesting trading strategies. If an LLM  is asked to analyze a news article from within its training window, the LLM already ``knows'' something about events that followed the news article, which taints the model's forecasts. Indeed, in other contexts, LLMs have been found to memorize training data that can later influence the output they produce \citep{NEURIPS2022_fa0509f4, 274574, steed-etal-2022-upstream}. 

Retraining an LLM on a rolling window would solve the problem of separating training and test data, but this is computationally infeasible and is likely to remain so for the foreseeable future. Researchers seeking to test LLM predictions have therefore been limited to the out-of-sample period that begins at the end of the model's training period. For ChatGPT, the training period runs through September 2021, leaving only the period since October 2021 for out-of-sample testing. (The engine behind ChatGPT is GPT-3.5-Turbo, which we refer to henceforth as GPT-3.5.)

To address this problem, we propose a simple tool: anonymizing company information in news text. This tool allows us to measure and mitigate the bias in testing an LLM's response to text from within its training window. To measure the bias, we compare the in-sample performance of trading strategies that use the original and anonymized text. We also compare the out-of-sample performance of the strategies and find evidence that anonymization is potentially useful even outside the training window.

\begin{figure}[htbp]
    \centering
    \subfloat[\textbf{The potential impact of look-ahead bias and distraction on model performance due to preexisting knowledge through training data.} The direction in which the bias will influence the model's prediction is unclear, as this is contingent on precisely how much information the LLM has memorized.]{{ \includegraphics[width=13cm,height=6cm]{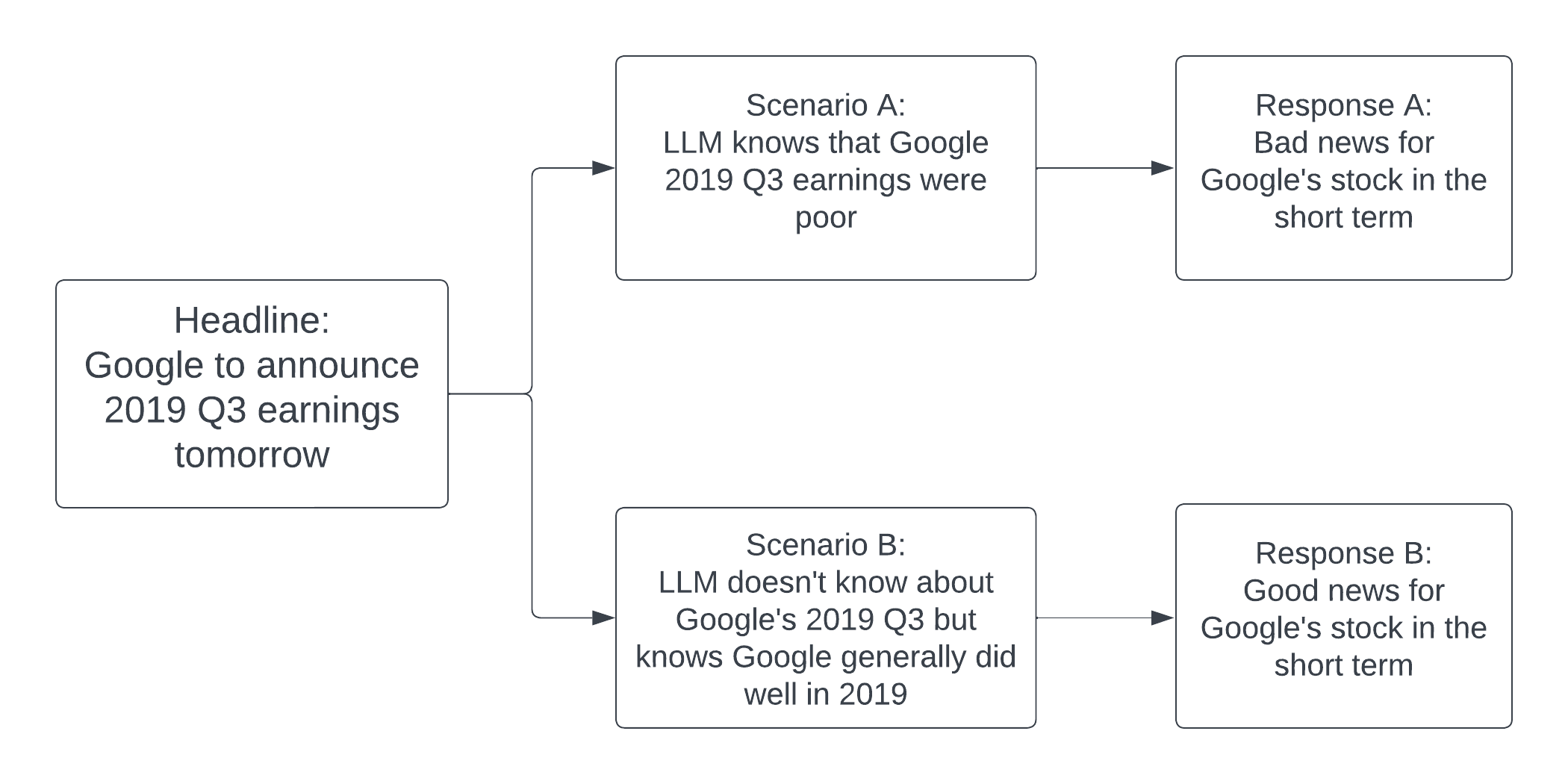} }}%
    \quad
    \subfloat[\textbf{Eliminating look-ahead bias and distraction through entity anonymization.} Because the actual company name has been removed and replaced with a random string, but the actual content of the headline has not been altered at all, the LLM is forced to process this replaced headline exclusively by analyzing the headline's language.]{{\includegraphics[width=13.5cm,height=2.2cm]{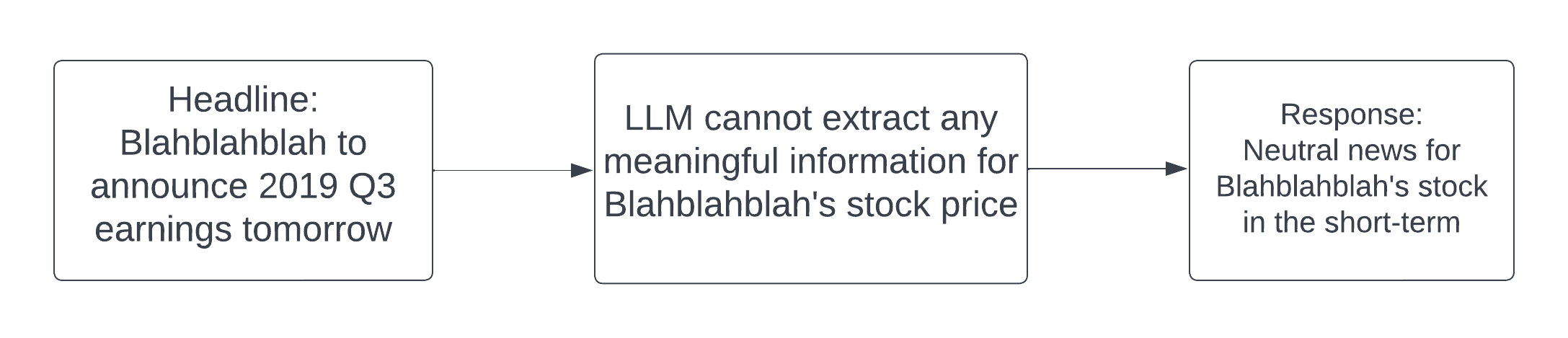} }}%
  \caption{\textbf{A scenario where look-ahead bias could potentially impact stock return predictions, and how it can be avoided through entity anonymization.}}%
  \label{fig: diagram}
\end{figure}

We consider trading strategies that buy a stock following good news about the stock and sell the stock following bad news. An LLM is used to judge whether news is good, bad, or neutral. In the example in Figure~\ref{fig: diagram}, an LLM is asked whether the headline, ``Google to announce 2019 Q3 earnings tomorrow'' is good or bad news for the company. The correct answer is clearly neither. But in the figure's Scenario A, the LLM knows that Google's 2019 Q3 earnings were poor and responds that the news is bad. We use the term \textit{look-ahead bias} to refer this type of scenario, in which the LLM has specific information about relevant events immediately following the news story.

In Scenario B, the LLM does not know specifically about Google's 2019 Q3 results, but it has general knowledge about Google, including the fact that Google, in general, did well in 2019. It therefore responds that the headline is good news for Google. We use the term \textit{distraction effect} to refer to the ways the LLM's general knowledge about a company might interfere with its analysis of the sentiment of news text. This interference may come from knowledge of the future, as in the example in the figure, but it could also come from past information about the company.

Part (b) of Figure~\ref{fig: diagram} shows how we mitigate both types of bias. We anonymize the headline, and the LLM then correctly responds that the news is neutral. Comparing results with and without anonymization allows us to quantify the overall bias in in-sample backtesting. The performance of de-biased trading strategies provides a better indication of out-of-sample performance.

Look-ahead bias should be positive: knowing the short-term response to news should produce higher returns than could be achieved in practice. The direction of the distraction effect is unclear: general knowledge about a company could be helpful or harmful in sentiment analysis. Also, whereas look-ahead bias is restricted to in-sample results, distraction is relevant out-of-sample as well. Indeed, we use ``distraction effect'' to capture all the ways an LLM's response to a news prompt differs from its response to an anonymized prompt, aside from  specific knowledge of events immediately following the news, which is captured by look-ahead bias.

%

We measure the performance of long-short trading strategies (based on the approach in \citet{lopezlira2023chatgpt}) that use GPT-3.5 for sentiment analysis, but we compare results using the original and replaced (anonymized) news headlines. To our surprise, we find that \emph{the replaced headlines generate higher in-sample returns}. Since look-ahead bias must be positive, we conclude that the distraction effect must be negative: the general knowledge GPT-3.5 has about companies does more harm than good in judging news sentiment in in-sample backtesting. This conclusion is robust in that it holds across two different news sources. The effect is particularly strong among larger companies, about which we expect GPT-3.5 to have greater general knowledge.

When we dig deeper into the drivers of these results, we find that GPT-3.5 suffers from a kind of overconfidence. We compare two scenarios --- one in which GPT-3.5 misjudges the sentiment in the original headline and judges the replaced headline as neutral, and one in which GPT-3.5 misjudges the sentiment in the replaced headline and judges the original headline as neutral. These are scenarios in which the distraction effect is, respectively, harmful and helpful. The losses when distraction is harmful are greater than the losses from failing to benefit from distraction when its effect is helpful. GPT-3.5 would do better by better recognizing what it does not know.

The performance comparison between the original and replaced headlines loses statistical significance out-of-sample, suggesting that most of the distraction effect we observe in-sample comes from knowledge of the future. However, for larger companies, we find borderline significant evidence that the replaced headlines outperform even out-of-sample. In other words, anonymizing news text is not just a tool for measuring in-sample bias; it may also be helpful in improving out-of-sample performance.

Section~\ref{s:methods} describes our data sources, trading strategies, and anonymization process. Section~\ref{s:results} presents our main findings, including in-sample and out-of-sample comparisons of trading strategies using original and replaced headlines; average returns conditional on original and replaced responses; the influence of company size; the predictive power of GPT scores; and a comparison of market betas. Section~\ref{s:discussion} summarizes our conclusions. Supplementary results are included in an appendix.

\section{Methods}
\label{s:methods}

\subsection{Data}

Our in-sample period begins in January 2015 and ends in September 2021, when GPT-3.5's training data ends \citep{openai}. Our out-of-sample period begins in October 2021 and ends in December 2022.


We use the Center for Research in Security Prices's (CRSP) datasets to obtain daily stock market data, specifically daily open and close prices, daily listed shares, and daily market cap. We use this information to calculate daily open-to-close returns and daily close-to-close returns for individual companies.


We use two sets of news headlines, which we refer to as the ``scraped'' and Thomson Reuters (or TR) datasets. To assemble our scraped collection, we use RavenPack as a guide. RavenPack is a popular provider of news from Dow Jones, but it prohibits uploading of its text to third parties, including GPT-3.5. We therefore search the web to find news headlines about specific companies on days RavenPack has news about those companies to assemble our own collection of headlines. We map each company's RavenPack ID to its CRSP security ID to match each company-headline pair with stock return data. This process yields 129,431 headlines over our sample period, representing 6,723 companies that were both mentioned in a RavenPack news item and included in the CRSP universe of stocks. The Dow Jones data underlying RavenPack covers a broad range of companies.

For our second news source we use the Thomson Reuters data assembled in \citet{glasserman2022} for S\&P 500 companies. That collection includes full news articles, but here we use only the headlines. Each article was previously linked to the CRSP ID of each S\&P 500 company mentioned in the article through a procedure described in \citet{glasserman2022}. This dataset yields 181,908 headlines representing 678 S\&P 500 companies over the sample period, all of which were both mentioned in at least one Thomson Reuters news headline and contained in the CRSP stock return dataset. The most important difference between the two collections is the median market cap (2.59 billion USD for scraped, 56.74 billion USD for Thomson Reuters). For a more extensive comparison of the Thomson Reuters and Dow Jones news sources, see \citet{glasserman2022}.


\subsection{GPT Usage}
\label{s:usage}

We follow the methodology of  \citet{lopezlira2023chatgpt} to perform sentiment analysis of news headlines using GPT-3.5. We use the following prompt as input to the language model:

\begin{quote}
    Forget all your previous instructions. Pretend you are a financial expert. You are a financial expert with stock recommendation experience. Answer “YES” if good
    news, “NO” if bad news, or “UNKNOWN” if uncertain in the first line. Then
    elaborate with one short and concise sentence on the next line. Is this headline
    good or bad for the stock price of [company name] in the short term? \\

    Headline: [headline]
\end{quote}

We use the GPT-3.5-Turbo version from OpenAI, which is the language model that powers ChatGPT. We set the model's temperature, a parameter that controls the determinism of the output, to 0 (most predictable) in order to maximize reproducibility of results.

Using the prompt, we ask the model to provide a recommendation and analysis for each headline, quantifying the results by mapping ``YES'' to 1, ``UNKNOWN'' to 0, and ``NO'' to -1. We then align each score with the appropriate market period, following \citet{lopezlira2023chatgpt}. 
For headlines that appear after 4 P.M. ET and before 6 A.M., the stock is bought at the next opening price and sold at the next closing price; for headlines after 6 A.M. and before 4 P.M., the stock is bought at the closing price and sold at the next day's close.
Scores are averaged if there are multiple headlines for a company within the same market period. 

\subsection{Trading Strategy}
\label{s:trading}


To evaluate the performance of the GPT recommendations and to analyze look-ahead bias, we examine three types of trading strategies advised by GPT scores: long-only, short-only, and long-short. The long-only strategy buys all companies with positive GPT scores and holds them for one day (open to close or close to close, as explained in Section~\ref{s:usage}). The short-only strategy short-sells all companies with negative GPT scores for one day. The long-short strategy does both. If a strategy does not have any corresponing recommendations within a given market period, it does not execute any trades. All portfolios are equal-weighted and rebalanced at the end of each market day. 


In more detail, the long-only strategy works as follows. At each market open, we identify all companies with a GPT score greater than 0 (corresponding to positive recommendations from news headlines) and calculate their average intraday market return over the subsequent day. At market close, we again identify all companies with a GPT score greater than 0 and calculate their average subsequent close-to-close market return.  We then calculate the total return for the day by computing the weighted average of the intraday market return and the overnight market return, weighted based on the number of companies contributing to each average. The short-only strategy works analogously, based on scores less than 0.

We stress that these strategies are not intended to be feasible; the in-sample versions clearly are not, and even out-of-sample we are ignoring transaction costs and short-sale constraints. We use these strategies as a mechanism for analyzing the de-biasing effect of anonymizing the input news text. 

\subsection{Headline Anonymization}

We explore the potential impact of look-ahead bias by first anonymizing the original headlines to hide the pertinent company name through a string replacement algorithm detailed below. Then, we repeat  the GPT processing and implement trading strategies with these revised GPT scores. 

First, we identify the official company name of the company that the headline discusses, which is available for all articles through RavenPack and our Thomson Reuters collection. We clean the official name by removing all punctuation, as well as common endings like ``Inc'', ``LLC'', or ``Limited'', which are unlikely to ever appear in standard news headlines. 

For each headline, we iterate through each substring with length equal to the length of the cleaned company name and compare it to the cleaned company name using fuzzy string matching. Specifically, we perform a set operation that isolates common tokens, then sort the strings, and finally paste the tokens back together. Next, we compute the normalized InDel similarity score between the two remaining strings, which is a score between 0 and 100, where 100 represents two identical strings. If the similarity score is greater than a threshold of 80, we identify a match between the substring and the company name. 

The InDel distance measures the number of insertions and deletions needed to transform one string into another.
In the case of company names, it is common for companies to be referred to through condensed versions of their official names, with certain words added or removed, making the InDel distance a suitable metric to fulfill our purpose. 

We then replace the substring with a random string, ``Blahblahblah'', thus anonymizing the actual company name.%
\footnote{We obtained similar results
using other random strings such as ``abcdefghijklmnopqrstuvwxyz'' or ''JREBEvphod'' instead of ``Blahblahblah''. Therefore, we continue to use ``Blahblahblah'' as our anonymized replacement string.} 
We then repeat this with all substrings of length less than the length of the company name to account for partial matches, such as to ensure that ``Costco'' is identified from the official name of ``Costco Wholesale Corporation''. In addition, for companies with names that are at least two words in length, we perform a case-sensitive search for an abbreviated version of the company's name (e.g. to find ``IBM'' from ``International Business Machine'' or ``GE" from ``General Electric''). 

We also observe that with some headlines, it may not be sufficient to conceal just the name of the company, as related information can be equally revealing of its identity. For instance, most LLMs will be able to identify that a replaced headline ``Blahblahblah announces release of new iPhone'' is relevant to Apple. This will be especially important for the Thomson Reuters dataset, which generally reports on larger companies that the LLM will possess more knowledge of.

To account for this, we make use of the Google Knowledge Graph, Google's public database that contains billions of facts about people, places, and things. For each company, we query the Knowledge Graph to extract the most relevant products and services, up to a limit of 20 (see Table \ref{tab: gkg} for example queries). We then search for each product or service in a given headline, and if found, replace it with the random string.
Table~\ref{tab:my_label} provides examples of the replacement process; note, for example, that ``Xbox'' is replaced along with ``Microsoft.''

\begin{table}[H]
    \centering
    \begin{tabular}{p{0.15\linewidth}  p{0.8\linewidth}}
    \toprule
    Query & Results \\
    \midrule
    \rowcolor[rgb]{ .875,  .89,  .898} Apple & [Apple, Apple, Apple Watch, AirPods, iPhone, iPad, Apples, AirPods Pro, iPhone 11, Apple TV] \\
    & \\
    Microsoft & [Microsoft Corporation, Microsoft 365, MSN, Microsoft Bing, OneDrive, Microsoft Word, Microsoft PowerPoint, Microsoft Office, Xbox, Microsoft Edge]\\
    & \\
    \rowcolor[rgb]{ .875,  .89,  .898} Amazon & [Amazon.com, Amazon Prime Video, Amazon Prime, Amazon Alexa, Amazon Music, Amazon Kindle, Amazon Fresh, Amazon Web Services, Amazon Echo, Amazon Luna] \\ 
    \bottomrule
    \end{tabular}
    \caption{\textbf{Examples of Google Knowledge Graph Queries.} 
    In a headline about Apple, we would anonymize all the terms related to Apple.}
    \label{tab: gkg}
\end{table}

We also tried using the Python library Spacy to perform named entity recognition, a natural language processing method that detects and categorizes important information in text, to identify company names. However, we observed that the library often failed to properly identify companies and organizations and frequently removed important content within the headline. Because such inaccuracies could significantly alter the meaning or substance of an article, we instead chose to develop the replacement algorithm detailed above. 

Our anonymization algorithm identifies and replaces the relevant company name or a relevant product or service in 100,401 out of 129,431 cases in the scraped headlines and 180,237 out of 181,908 cases in the Thomson Reuters headlines. Headlines that were not modified by the algorithm were processed as is in both sets of headlines, as they do not directly mention the relevant company and therefore can be used as textual input for both the original trading strategy and the strategy using replaced headlines.

Throughout the paper, we will refer to news headlines as either ``original'' or ``replaced'', where the latter indicates that the headline has been modified such that relevant company identifiers have been replaced and anonymized. Original headlines will produce ``original scores'' that advise ``original portfolios,'' and replaced headlines will produce ``replaced scores'' that generate ``replaced portfolios''. 
Our anonymization procedure yields separate sets of replaced and original headlines for the scraped data and the Thomson Reuters data.

\begin{table}[H]
    \centering
    \begin{tabular}{p{4cm}  p{5cm} p{5cm} }
        \toprule
        Official company name & Original headline & Replaced headline \\
        \midrule
         \rowcolor[rgb]{ .875,  .89,  .898}  \footnotesize Walgreens Boots Alliance Inc. & \footnotesize Walgreens 4Q Profit Falls as Overseas Unit Struggles Amid Pandemic & \footnotesize Blahblahblah 4Q Profit Falls as Overseas Unit Struggles Amid Pandemic \\
        \midrule
         \footnotesize Microsoft Corporation & \footnotesize `Halo Infinite' delayed, will not debut with new Xbox; Microsoft stock slips & \footnotesize 'Halo Infinite' delayed, will not debut with new Blahblahblah; Blahblahblah stock slips \\
        \midrule
         \rowcolor[rgb]{ .875,  .89,  .898}  \footnotesize Phillips Petroleum Company & \footnotesize Analysis: Market for U.S. oil acreage booms along with crude price recovery & \footnotesize Analysis: Market for U.S. oil acreage booms along with crude price recovery \\
        \midrule
        \footnotesize Advanced Micro Devices Inc. & \footnotesize AMD's Stock Sinks, Xilinx Soars After WSJ Report Of Advanced Merger Talks -- MarketWatch & \footnotesize Blahblahblah's Stock Sinks, Xilinx Soars After WSJ Report Of Advanced Merger Talks -- MarketWatch \\
        \midrule
         \rowcolor[rgb]{ .875,  .89,  .898}  \footnotesize Xilinx Inc. & \footnotesize AMD's Stock Sinks, Xilinx Soars After WSJ Report Of Advanced Merger Talks -- MarketWatch & \footnotesize AMD's Stock Sinks, Blahblahblah Soars After WSJ Report Of Advanced Merger Talks -- MarketWatch \\
        \bottomrule
    \end{tabular}
    \caption{\textbf{Select samples of the results of anonymization algorithm.} The algorithm extracts and replaces words and phrases that can identify a company, such as its name, abbreviation, or products. In the first row, just the company name is replaced, and in the second example, the company name and a relevant product are replaced. In the third example, the original and replaced headlines are identical because the company is not directly mentioned, while in the fourth example, the company's abbreviation is identified and replaced. The fifth example, a repeat of the fourth but with a different company to be identified, shows that the replaced strings are contingent on the company 
    to be named in the GPT prompt.}
    \label{tab:my_label}
\end{table}




\section{Results}
\label{s:results}

%

We now compare the performance of trading strategies using original and replaced headlines, conducting the comparison separately in-sample and out-of-sample and separately for each news source. We then drill down to compare returns conditional on various combinations of responses from GPT-3.5 to the original and replaced headlines. We investigate how company size influences the response to the original and replaced headlines. We compare the predictive value of sentiment scores based on the two types of headlines, and we compare market betas for original and replaced portfolios.

\subsection{Trading Performance}

\begin{figure}[H]
    \centering
    {{ \includegraphics[width=10cm,height=7cm]{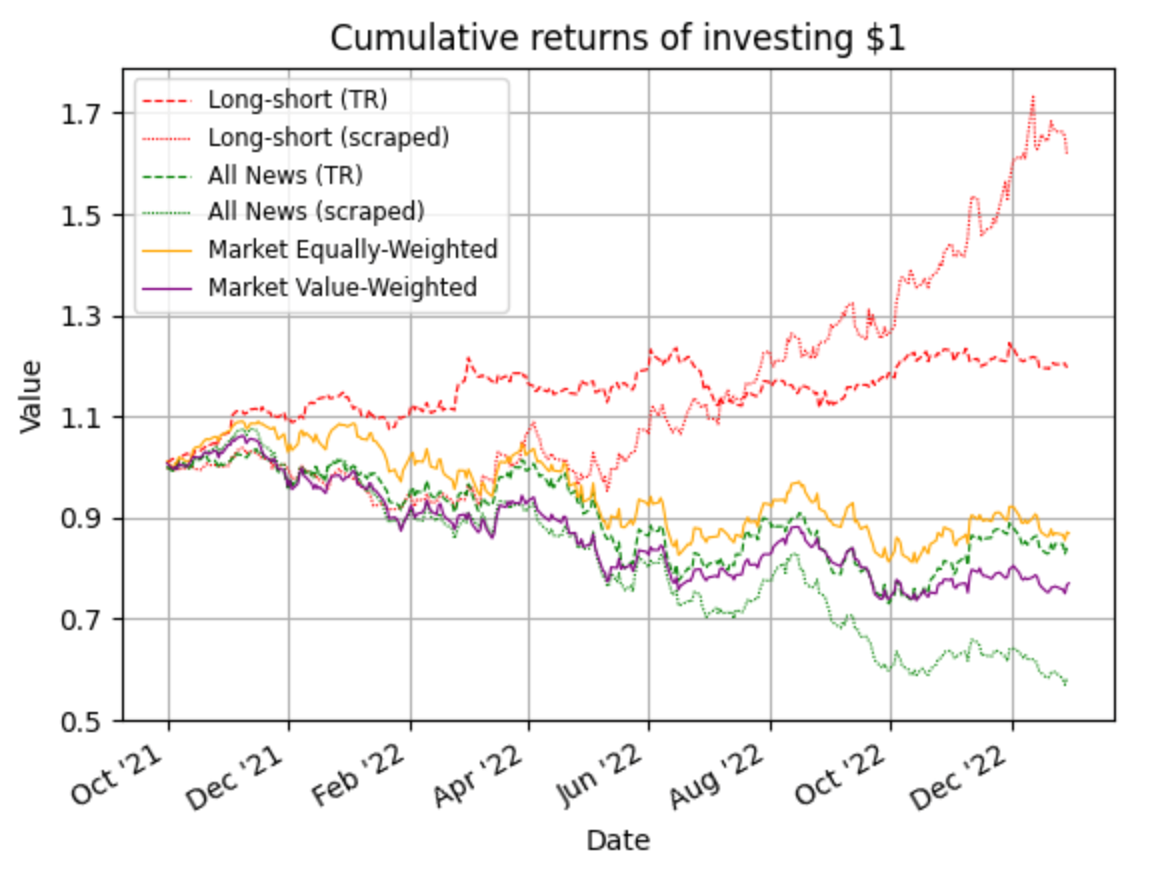} }}%
  \caption{\textbf{Out-of-sample trading performance using the original scraped and original TR headlines.}
The dotted red line and the dashed red line show cumulative returns for the long-short strategy using original scraped and original TR headlines, respectively. The dotted green line and the dashed green line show cumulative returns from holding all companies with any news on the previous day, based on the scraped and TR headlines, respectively.  For comparison, the yellow and purple lines show equal-weighted and value-weighted cumulative returns to the overall market.}%
  \label{fig: returns}
\end{figure}

We begin by confirming the out-of-sample profitability of trading on GPT-3.5's sentiment scores. Figure \ref{fig: returns} shows that the long-short trading strategies implemented using the original scraped and original TR headlines as input and GPT-3.5's output substantially outperform market returns over the out-of-sample period. The results in in the figure generally align with \citet{lopezlira2023chatgpt}, although \citet{lopezlira2023chatgpt} report even higher long-short returns. We are using a simplified version of their trading strategy (and not using RavenPack's relevance score, for example) because our objective is to measure and mitigate in-sample bias. A simpler strategy with less fine-tuning provides greater transparency for the effect our de-biasing approach. In addition, our headline collections differ from the headlines used in \citet{lopezlira2023chatgpt}.

In Figure \ref{fig: returns}, trading on the scraped headlines generates higher returns than trading on the TR headlines. This may be due to a greater supply of bad-news stories in the scraped dataset, which provide short-selling opportunities: the scraped ``All News'' portfolio substantially underperforms the market in Figure \ref{fig: returns}. The TR ``All News'' portfolio is roughly in line with the market, suggesting that the TR headlines provide fewer profitable signals during the out-of-sample period. We will have more to say about the news sources in subsequent sections, and we examine returns on the long and short sides separately in the appendix.

\subsection{Comparing Daily Average Portfolio Returns}
\label{s:comparing}


We now turn to comparisons of the average daily returns using original and replaced headlines. We compare results for these two strategies separately during the in-sample and out-of-sample periods and separately for each of our two news sources.

Table \ref{tab:sig tests} summarizes our results. The top panel compares the original and replaced strategies during the in-sample period. For both news sources \emph{the replaced headlines generate higher average returns than the original headlines.} (Specifically, we see an outperformance of $30.97-25.08=5.89$ basis points per day using the scraped headlines and $81.64-77.27=4.37$ using the TR headlines.) The difference between the replaced and original mean returns is statistically significant using the TR data ($p=0.017$) and borderline significant ($p=0.066$) using the scraped data because of the higher volatility of those returns.

\begin{table}[htbp]
  \centering
    \begin{tabular}{clccccc}
    \toprule
          &       & \multicolumn{2}{c}{Scraped} &       & \multicolumn{2}{c}{Thomson Reuters} \\
          &       & Original & Replaced &       & Original & Replaced \\
    \midrule
    \multirow{5}[4]{*}{In-sample} & \cellcolor[rgb]{ .875,  .89,  .898}No. of obs. & \cellcolor[rgb]{ .875,  .89,  .898}1699 & \cellcolor[rgb]{ .875,  .89,  .898}1699 & \cellcolor[rgb]{ .875,  .89,  .898} & \cellcolor[rgb]{ .875,  .89,  .898}1695 & \cellcolor[rgb]{ .875,  .89,  .898}1695 \\
          & Mean  & 25.08 & 30.97 &       & 10.74 & 13.84 \\
          & \cellcolor[rgb]{ .875,  .89,  .898}Std. dev. & \cellcolor[rgb]{ .875,  .89,  .898}183.87 & \cellcolor[rgb]{ .875,  .89,  .898}219.32 & \cellcolor[rgb]{ .875,  .89,  .898} & \cellcolor[rgb]{ .875,  .89,  .898}77.27 & \cellcolor[rgb]{ .875,  .89,  .898}81.64 \\
\cmidrule{2-7}          & t-stat & \multicolumn{2}{c}{-1.84} &       & \multicolumn{2}{c}{-2.38} \\
          & \cellcolor[rgb]{ .875,  .89,  .898}p-value & \multicolumn{2}{c}{\cellcolor[rgb]{ .875,  .89,  .898}0.066} & \cellcolor[rgb]{ .875,  .89,  .898} & \multicolumn{2}{c}{\cellcolor[rgb]{ .875,  .89,  .898}0.017} \\
    \midrule
    \multirow{5}[4]{*}{Out-of-sample} & No. of obs. & 314   & 314   &       & 314   & 314 \\
          & \cellcolor[rgb]{ .875,  .89,  .898}Mean & \cellcolor[rgb]{ .875,  .89,  .898}16.32 & \cellcolor[rgb]{ .875,  .89,  .898}11.09 & \cellcolor[rgb]{ .875,  .89,  .898} & \cellcolor[rgb]{ .875,  .89,  .898}6.07 & \cellcolor[rgb]{ .875,  .89,  .898}12.23 \\
          & Std. dev. & 141.52 & 148.26 &       & 85.25 & 98.21 \\
\cmidrule{2-7}          & \cellcolor[rgb]{ .875,  .89,  .898}t-stat & \multicolumn{2}{c}{\cellcolor[rgb]{ .875,  .89,  .898}1.20} & \cellcolor[rgb]{ .875,  .89,  .898} & \multicolumn{2}{c}{\cellcolor[rgb]{ .875,  .89,  .898}-1.86} \\
          & p-value & \multicolumn{2}{c}{0.23} &       & \multicolumn{2}{c}{0.064} \\
    \bottomrule
    \end{tabular}%
    \caption{\textbf{Summary statistics and significance tests comparing long-short portfolios advised by original headlines and replaced headlines.} Each observation represents the average return of the portfolio on a given trading day. P-values are calculated from paired t-tests, with returns on the same day paired. Each t-stat tests the difference in the means immediately above it. For analysis with long-only portfolios and short-only portfolios, see the appendix.}
  \label{tab:sig tests}%
\end{table}%


The outperformance by the replaced headlines is surprising --- we would expect that the strategy using the original headlines would benefit from look-ahead bias in its in-sample performance. The fact that the original headlines underperform indicates that GPT's general knowledge about companies --- the distraction effect --- negatively affects its ability to evaluate news sentiment and outweighs any positive look-ahead bias. We will examine this effect in greater detail in the next sections.

The lower panel of Table \ref{tab:sig tests} compares the performance of original and replaced headlines in the out-of-sample period. Using the scraped data, the difference in means changes sign but the difference is not statistically significant. The comparison using the TR data suggests that the negative distraction effect persists out-of-sample, although the difference is now only borderline significant. We will return to this contrast between the two news sources in Section~\ref{s:mcap}.

\subsection{Performance of Replaced Headlines}

\subsubsection{Illustrative Examples}

To better illustrate the phenomenon suggested by Table \ref{tab:sig tests}, we analyze some specific examples of how entity anonymization can affect the output that GPT-3.5 produces.

\begin{table}[H]
  \centering
  
    \begin{tabular}{p{4.3em}p{4.4em}p{9.9em}p{9.9em}p{4.9em}p{4.7em}}
    \toprule
    Firm & Headline & Original Response & Replaced Response & Short-term & Long-term \\
    \midrule
    \rowcolor[rgb]{ .875,  .89,  .898} Facebook & \small Facebook not planning to extend European privacy law globally & \small YES - This statement is good for the price of Facebook in the short term as it indicates that the company will not face additional regulatory challenges and potential fines related to extending European privacy laws globally. & \small NO - The statement is bad for the price of Blahblahblah in the short term as it suggests that the company may face regulatory challenges and potential limitations in expanding its user base globally. & Stock down 0.6\% & Stock up 25\% over quarter due to greater worldwide usage  \\
     Allergan & \small Pfizer to buy Allergan in 160 billion deal & \small NO - The announcement of Pfizer's acquisition of Allergan in a 160 billion deal is bad news for the price of Allergan in the short term as it may lead to increased debt and integration challenges. & \small YES - This news is good for the price of Blahblahblah in the short term as the acquisition by Pfizer in a significant deal indicates potential growth and expansion for the company. & Stock up 6\% & Deal failed 6 months later, stock down 15\% \\
    \bottomrule
    \end{tabular}%
    \caption{\textbf{Select examples where entity anonymization significantly affects the output that GPT-3.5 produces.} The original response is GPT-3.5's response when provided the headline as is. The replaced response is GPT-3.5's response when provided the headline with the relevant company name replaced with a random string. The short-term outcome is the immediate movement of the company's stock in the next trading period (i.e. the return that the trading strategies advised by GPT output would attain). The long-term outcome in the long-term impact of the given headline on the company's stock.}
    \label{tab: case study}
\end{table}%

The first example of Table \ref{tab: case study} relates to Facebook (now Meta) through a headline reporting on their decision to not extend European privacy laws internationally. GPT-3.5 interprets this headline as good news for Facebook's stock, as it predicts that fewer privacy laws globally will reduce regulatory challenges. When the headline is modified such that the word Facebook is replaced with a random string, GPT interprets the headline as bad news due to potentially reduced user base as a result of fewer privacy laws. In reality, following the headline, the stock of Facebook decreased by 0.6\%, meaning that the replaced response produced the correct investment decision. However, in the longer term, throughout the remainder of the quarter, the stock of Facebook increased by over 25\% due to greater worldwide adoption. While it is difficult to determine precisely what influences the LLM to make its decisions, given that the only modification to the headline between these two responses was the anonymization of Facebook, it appears likely that GPT-3.5—when processing the original headline—leverages some type of existing knowledge of Facebook's stock performance rather than solely its language processing capabilities when making this prediction.

In the second example, the headline reports that pharmaceutical company Pfizer is set to buy its counterpart Allergan in a 160 billion dollar deal. Given the original headline, GPT-3.5's response is that the news is bad for Allergan's stock, as the acquisition may cause increased debt and integration challenges. With the anonymized headline, the replaced response is that the headline is good news because it indicates potential growth and expansion for the company. Similarly, given that the only change to the headline was the presence or absence of Allergan's name, it seems likely that GPT-3.5 attempted to utilize its preexisting knowledge of Allergan to generate a prediction given the original headline. Indeed, while Allergan's stock significantly increased immediately following the headline—making the replaced response again correct—the deal ultimately fell through 6 months later, causing the stock of Allergan to plunge 15\%. Given the size and scale of this attempted acquisiton in the pharma industry, 
GPT-3.5 may have gained knowledge of the event and its aftermath through its training data, which then influenced or interfered with its evaluation of the headline's sentiment.

\subsubsection{Classification Analysis Based on Response Accuracy}

We now argue that the in-sample outperformance of the replaced strategy is due, in part, to a type of ``overconfidence'' in GPT-3.5: some of the largest losses affecting the original strategy occur in scenarios where GPT would provide a neutral response to the replaced headline, but it responds positively or negatively to the original headline.

To make this type of comparison, we classify
every observation into one of nine categories, depending on the ``correctness'' of GPT-3.5's original response and replaced response. A response is deemed correct if the recommendation aligns with the direction of the daily stock return; for instance, a GPT score of 1 is correct if the relevant company experiences positive returns on the next trading day. 

\begin{table}[H]
    \centering
    \begin{tabular}{p{3cm} p{9cm}}
    \toprule
    Category & Definition \\
    \midrule
    \rowcolor[rgb]{ .875,  .89,  .898} \small orig \checkmark & \small The original GPT score is correct \\
    \small orig $\varnothing$ & \small The original GPT score is zero \\
    \rowcolor[rgb]{ .875,  .89,  .898} \small orig X & \small The original GPT score is incorrect \\
     \small rep \checkmark & \small The replaced GPT score is correct \\
    \rowcolor[rgb]{ .875,  .89,  .898} \small rep $\varnothing$ & \small The replaced GPT score is zero \\
    \small rep X & \small The replaced GPT score is incorrect \\
    \bottomrule
    \end{tabular}
    \caption{\textbf{The classifications that place each observation into one of nine categories, depending on the correctness of its original GPT score and its replaced GPT score.}}
\end{table}

We begin by dividing the headlines into the in-sample period and the out-of-sample period. We then categorize every observation. For each category, we compute an original return (the average return of a long-short portfolio generated with the original GPT scores of the observations in the given subset) and a replaced return (average return of a long-short portfolio generated with the replaced GPT scores of the observations in the given subset). Additionally, we document the total proportion of observations within each category.

Table \ref{tab: scrap_class} summarizes this analysis. We focus on the in-sample comparison. From the table we see that the replaced headlines produce the wrong answer 38.95\% ($=2.59\% + 20.48\% + 15.88\%$) of the time whereas the original headlines are wrong only 17.6\% ($=0.02\%+1.7\% + 15.88\%$) of the time. But when the original response is wrong and the replaced response is neutral, the original strategy loses 242 basis points, on average. In contrast, when the replaced response is wrong and the original response is neutral, the replaced strategy loses only 97 basis points, on average. Thus, although the original response is wrong less often, when it is wrong (and the replaced response remains neutral) it suffers comparatively large losses. The original headlines lead GPT to make some very bad trades that the replaced strategy sits out. We see a similar pattern out-of-sample, where we can compare the losses of 289 basis points and 130 basis points in the corresponding scenarios.

\begin{table}[htbp]
  \centering
    \begin{tabular}{ccc|ccc|ccc|ccc}
    \toprule
          \multicolumn{3}{c}{}       & \multicolumn{3}{c}{rep \checkmark} & \multicolumn{3}{c}{rep $\varnothing$} & \multicolumn{3}{c}{rep X} \\
          &       &       & orig\_ret & rep\_ret & prop  & orig\_ret & rep\_ret & prop  & orig\_ret & rep\_ret & prop \\
    \midrule
    \multicolumn{1}{c}{\multirow{3}[2]{*}{\rotatebox{90}{in}}} & \cellcolor[rgb]{ .875,  .89,  .898}orig \checkmark & \cellcolor[rgb]{ .875,  .89,  .898} & \cellcolor[rgb]{ .875,  .89,  .898}318 & \cellcolor[rgb]{ .875,  .89,  .898}318 & \cellcolor[rgb]{ .875,  .89,  .898}15.11\% & \cellcolor[rgb]{ .875,  .89,  .898}269 & \cellcolor[rgb]{ .875,  .89,  .898}0 & \cellcolor[rgb]{ .875,  .89,  .898}1.69\% & \cellcolor[rgb]{ .875,  .89,  .898}278 & \cellcolor[rgb]{ .875,  .89,  .898}-4 & \cellcolor[rgb]{ .875,  .89,  .898}2.59\% \\
          & orig $\varnothing$ &       & 0     & 350   & 1.27\% & 0     & 0     & 41.26\% & 0     & -97   & 20.48\% \\
          & \cellcolor[rgb]{ .875,  .89,  .898}orig X & \cellcolor[rgb]{ .875,  .89,  .898} & \cellcolor[rgb]{ .875,  .89,  .898}-208 & \cellcolor[rgb]{ .875,  .89,  .898}209 & \cellcolor[rgb]{ .875,  .89,  .898}0.02\% & \cellcolor[rgb]{ .875,  .89,  .898}-242 & \cellcolor[rgb]{ .875,  .89,  .898}0 & \cellcolor[rgb]{ .875,  .89,  .898}1.70\% & \cellcolor[rgb]{ .875,  .89,  .898}-268 & \cellcolor[rgb]{ .875,  .89,  .898}-267 & \cellcolor[rgb]{ .875,  .89,  .898}15.88\% \\
    \midrule
    \multicolumn{1}{c}{\multirow{3}[2]{*}{\rotatebox{90}{out}}} & orig \checkmark &       & 313   & 313   & 20.42\% & 319   & 0     & 1.81\% & 283   & -5    & 2.08\% \\
          & \cellcolor[rgb]{ .875,  .89,  .898}orig $\varnothing$ & \cellcolor[rgb]{ .875,  .89,  .898} & \cellcolor[rgb]{ .875,  .89,  .898}0 & \cellcolor[rgb]{ .875,  .89,  .898}224 & \cellcolor[rgb]{ .875,  .89,  .898}1.52\% & \cellcolor[rgb]{ .875,  .89,  .898}0 & \cellcolor[rgb]{ .875,  .89,  .898}0 & \cellcolor[rgb]{ .875,  .89,  .898}37.51\% & \cellcolor[rgb]{ .875,  .89,  .898}0 & \cellcolor[rgb]{ .875,  .89,  .898}-130 & \cellcolor[rgb]{ .875,  .89,  .898}13.50\% \\
          & orig X &       & -658  & 670   & 0.01\% & -289  & 0     & 1.92\% & -290  & -290  & 21.23\% \\
    \bottomrule
    \end{tabular}%
  \caption{\textbf{Classification analysis of scraped headlines.} The orig\_ret and rep\_ret columns refer to the original return and replaced return, measured in basis points, and the prop column refers to the proportion of observations in each category.}
  \label{tab: scrap_class}
\end{table}%

Table \ref{tab: comp} reports the corresponding results using the TR headlines. The results are similar but show an even greater advantage to the replaced headlines. In-sample, the original response is now incorrect in over 16\% of observations, while the replaced response is incorrect in fewer than 13\% of observations. We again see an ``overconfidence'' effect: when the original response is wrong and the replaced response is neutral, the original strategy loses 138 basis points; the reverse scenario has an average loss of only 91 basis points. A similar comparison holds out-of-sample.

The results of Tables~\ref{tab: scrap_class} and~\ref{tab: comp} suggest that the types of errors shown in Table \ref{tab: case study} occur on a broader scale.

\begin{table}[htbp]
  \centering
    \begin{tabular}{cc|ccc|ccc|cccc}
    \toprule
          \multicolumn{2}{c}{}    & \multicolumn{3}{c}{rep \checkmark} & \multicolumn{3}{c}{rep  $\varnothing$} & \multicolumn{3}{c}{rep X} \\
          &    & \multicolumn{1}{c}{orig\_ret} & \multicolumn{1}{c}{rep\_ret} & \multicolumn{1}{c|}{prop} & \multicolumn{1}{c}{orig\_ret} & \multicolumn{1}{c}{rep\_ret} & \multicolumn{1}{c}{prop} & \multicolumn{1}{|c}{orig\_ret} & \multicolumn{1}{c}{rep\_ret} & \multicolumn{1}{c}{prop} \\
    \midrule
    \multicolumn{1}{c}{\multirow{3}[2]{*}{\rotatebox{90}{in}}} & \cellcolor[rgb]{ .875,  .89,  .898}orig \checkmark & \cellcolor[rgb]{ .875,  .89,  .898}185 & \cellcolor[rgb]{ .875,  .89,  .898}185 & \cellcolor[rgb]{ .875,  .89,  .898}11.70\% & \cellcolor[rgb]{ .875,  .89,  .898}141 & \cellcolor[rgb]{ .875,  .89,  .898}0 & \cellcolor[rgb]{ .875,  .89,  .898}5.17\% & \cellcolor[rgb]{ .875,  .89,  .898}24 & \cellcolor[rgb]{ .875,  .89,  .898}-10 & \cellcolor[rgb]{ .875,  .89,  .898}0.20\% \\
          & orig  $\varnothing$ &       0     & 98    & 0.97\% & 0     & 0     & 64.19\% & 0     & -91   & 1.61\% \\
          & \cellcolor[rgb]{ .875,  .89,  .898}orig X & \cellcolor[rgb]{ .875,  .89,  .898}-16 & \cellcolor[rgb]{ .875,  .89,  .898}16 & \cellcolor[rgb]{ .875,  .89,  .898}0.09\% & \cellcolor[rgb]{ .875,  .89,  .898}-138 & \cellcolor[rgb]{ .875,  .89,  .898}0 & \cellcolor[rgb]{ .875,  .89,  .898}4.99\% & \cellcolor[rgb]{ .875,  .89,  .898}-164 & \cellcolor[rgb]{ .875,  .89,  .898}-164 & \cellcolor[rgb]{ .875,  .89,  .898}11.10\% \\
    \midrule
    \multicolumn{1}{c}{\multirow{3}[2]{*}{\rotatebox{90}{out}}} & orig \checkmark &       222   & 222   & 13.88\% & 187   & 0     & 4.62\% & 27    & -6    & 0.25\% \\
          & \cellcolor[rgb]{ .875,  .89,  .898}orig  $\varnothing$ & \cellcolor[rgb]{ .875,  .89,  .898}0 & \cellcolor[rgb]{ .875,  .89,  .898}118 & \cellcolor[rgb]{ .875,  .89,  .898}0.95\% & \cellcolor[rgb]{ .875,  .89,  .898}0 & \cellcolor[rgb]{ .875,  .89,  .898}0 & \cellcolor[rgb]{ .875,  .89,  .898}62.60\% & \cellcolor[rgb]{ .875,  .89,  .898}0 & \cellcolor[rgb]{ .875,  .89,  .898}-101 & \cellcolor[rgb]{ .875,  .89,  .898}1.53\% \\
          & orig X &   -11   & 11    & 0.07\% & -176  & 0     & 2.66\% & -213  & -213  & 13.39\% \\ 
    \bottomrule
    \end{tabular}%
    \caption{\textbf{Classification analysis of Thomson Reuters headlines.} The orig\_ret and rep\_ret columns refer to the original return and replaced return, measured in basis points, and the prop column refers to the proportion of observations in each category.} 
    \label{tab: comp}
    
\end{table}%

The analysis in Tables~\ref{tab: scrap_class} and~\ref{tab: comp} is based on the sentiment scores of individual headlines. Our trading strategy averages scores for all headlines about a given company each market period, and the weight given to each company depends on the total number of companies with news that day. The mean returns in Tables~\ref{tab: scrap_class} and~\ref{tab: comp} 
are therefore not directly comparable to the mean strategy returns in Table~\ref{tab:sig tests}. Tables~\ref{tab: scrap_class} and~\ref{tab: comp}  allow us to focus more directly on the classification errors in individual headlines, while the analysis performed in Table 3 in conducted on the portfolio level.

\subsection{Market Cap of Recommended Companies}
\label{s:mcap}


We expect GPT-3.5 to be exposed to more text about larger companies in its training and therefore to be more likely to return positive or negative responses to news headlines about these companies. With this idea in mind, we examine the market cap of companies  for which GPT-3.5 makes buy or sell recommendations. We do this using the scraped headlines because our TR data is already restricted to S\&P 500 companies.

\begin{table}[htbp]
  \centering
    \begin{tabular}{c|lrrrrr}
    \toprule
    \multicolumn{1}{r}{} &       & \multicolumn{1}{l}{long} & \multicolumn{1}{l}{short} & \multicolumn{1}{l}{other} & \multicolumn{1}{l}{long - other} & \multicolumn{1}{l}{short - other} \\
    \midrule
    \multirow{3}[2]{*}{in-sample} & \cellcolor[rgb]{ .875,  .89,  .898}Mean & \cellcolor[rgb]{ .875,  .89,  .898}35.37 & \cellcolor[rgb]{ .875,  .89,  .898}42.37 & \cellcolor[rgb]{ .875,  .89,  .898}17.91 & \cellcolor[rgb]{ .875,  .89,  .898}17.46*** & \cellcolor[rgb]{ .875,  .89,  .898}24.46*** \\
          & Std. dev. & 129.17 & 139.27 & 93.22 &       &  \\
    \multicolumn{1}{c}{} & \cellcolor[rgb]{ .875,  .89,  .898}No. of obs. & \cellcolor[rgb]{ .875,  .89,  .898}27454 & \cellcolor[rgb]{ .875,  .89,  .898}9473 & \cellcolor[rgb]{ .875,  .89,  .898}62898 & \cellcolor[rgb]{ .875,  .89,  .898} & \cellcolor[rgb]{ .875,  .89,  .898} \\
    \midrule
    \multirow{3}[2]{*}{out-of-sample} & Mean  & 46.30  & 70.32 & 32.18 & 14.12*** & 38.15*** \\
          & \cellcolor[rgb]{ .875,  .89,  .898}Std. dev. & \cellcolor[rgb]{ .875,  .89,  .898}143.74 & \cellcolor[rgb]{ .875,  .89,  .898}217.04 & \cellcolor[rgb]{ .875,  .89,  .898}123.47 & \cellcolor[rgb]{ .875,  .89,  .898} & \cellcolor[rgb]{ .875,  .89,  .898} \\
          & No. of obs. & 9602  & 4452  & 15552 &       &  \\
    \bottomrule
    \multicolumn{5}{l}{\cellcolor{white}\footnotesize* p $<$ 0.10, ** p $<$ 0.05, *** p $<$ 0.01} \\
    \end{tabular}%
  \caption{\textbf{Comparison of the average market cap of companies recommended by GPT-3.5 to other companies with news, using scraped headlines.} Each observation is the market cap of a company within the given category, as measured in billions of USD. The ``long'' and ``short'' categories represent the average market cap of companies that GPT-3.5 recommended to buy or recommended to short-sell, respectively. The ``other'' category represents the average market cap of all other companies with news.}
  \label{tab:mcap}%
\end{table}%

Table \ref{tab:mcap} compares the average market cap of companies in the long-only and short-only portfolios and all other stocks mentioned in our scraped headlines. 
The table shows that, indeed, the companies traded by GPT-3.5 are significantly larger than the average company with news. In-sample, this pattern is consistent with both look-ahead bias and a distraction effect. But the pattern persists out-of-sample, where it cannot be explained by look-ahead bias and is thus more likely due to a distraction effect: GPT is more likely to provide a positive or negative recommendation when it knows more about a company, even in the absence of specific information relevant to a headline, and we expect it to know more about larger companies.

This tendency may also explain why, in out-of-sample trading on TR headlines, the replaced strategy outperforms the original strategy by a borderline significant amount; see Table~\ref{tab:sig tests}. The stocks covered by our TR headlines are all S\&P 500 stocks and are therefore larger, on average, than most other stocks. If GPT-3.5 tends to be misled by exogenous information, then anonymizing headlines should be most effective for larger stocks.

\subsection{Predictive Power of GPT Scores}



Section~\ref{s:comparing} compared original and replaced scores through trading strategies. We can also measure the predictive power of the scores directly by regressing returns on the scores.
Our basic specification takes the form
\begin{equation}
    r_{i, t+1} = a_i + b_t + \beta x_{i,t} + \epsilon_{i, t+1},
    \label{rspec}
\end{equation}
where  $r_{i, t+1}$ is the single-period return of firm $i$ from $t$ to $t+1$;
$a_i$ and $b_t$ are firm-fixed and time-fixed effects, respectively; and $x_{i,t}$ is the GPT score for firm $i$ in period $t$. Recall that the return $r_{i,t+1}$ is either an intraday return (in which case $x_{i,t}$ is calculated from news in the prior 4 P.M. to 6 A.M. period) or a close-to-close return (in which case $x_{i,t}$ is calculated from news in the prior 6 A.M. to 4 P.M. period).

The score $x_{i,t}$ may be based on original or replaced headlines, resulting in respective coefficients $\beta_1$ and $\beta_2$. We want to compare these coefficients and, in particular, test the hypothesis
\begin{equation}
    H_0: \beta_1 = \beta_2. 
    \label{null}
\end{equation}
The original and replaced scores are on the same scale, so comparing the coefficients is meaningful.

To test (\ref{null}), we set up two versions of equation (\ref{rspec})—one using original headlines and one using replaced headlines—as a system of seemingly unrelated equations to estimate them jointly. 
We stack $r_{i, t+1} $ on top of itself, include both sentiment scores as independent variables, and pad the remaining values with 0 to form the following equation: 
\begin{equation}
    \begin{pmatrix} r_{i, t+1} \\ r_{i, t+1} \end{pmatrix} = \begin{pmatrix}
        x_{1,t} & 0 \\ 0 & x_{2,t}
    \end{pmatrix} \begin{pmatrix}
        \beta_1 \\ \beta_2
    \end{pmatrix} + \begin{pmatrix}
        a_{1,i} \\ a_{2,i}
    \end{pmatrix} + \begin{pmatrix}
        b_{1,t} \\ b_{2,t}
    \end{pmatrix} + \begin{pmatrix}
        e_{1,i, t+1} \\ e_{2,i, t+1}
    \end{pmatrix}
    \label{eq: stacked1}
\end{equation}
This regression estimates the two coefficients jointly and allows us to test (\ref{null}) 
using a Wald chi-squared test.

Table~\ref{tab: regression2} reports results for four regressions of the form in (\ref{eq: stacked1}), using scraped and TR headlines, in-sample and out-of-sample. Each column reports the coefficients for the original and replaced scores. The Wald test and the p-value at the bottom of each column test equality of the coefficients for original and replaced scores.

In each of the four columns, the replaced-score coefficient is at least as large as the original-score coefficient, indicating that anonymizing the headlines increases (or does not decrease) the predictive value of the scores. Using the scraped headlines, the difference between the original and replaced coefficients is not statistically significant, but the magnitude of the difference is greater in-sample than out-of-sample.
Using the TR headlines, the difference is highly significant in-sample and borderline significant out-of-sample. These findings are consistent with patterns discussed previously: the distraction effect appears to be stronger than the look-ahead bias, particularly among the larger companies covered by the TR data; the distraction effect can persist even out-of-sample, indicating that anonymization can be helpful even for out-of-sample performance.

\begin{table}[htbp]
  \centering
    \begin{tabular}{lccccc}
    \toprule
          & \multicolumn{2}{c}{Scraped} &       & \multicolumn{2}{c}{Thomson-Reuters} \\
    \midrule
          & In-Sample & Out-of-Sample &       & In-Sample & Out-of-Sample \\
    \midrule
    \rowcolor[rgb]{ .875,  .89,  .898} orig\_score & 45.9*** & 23.5*** &       & 15.7*** & 13.4*** \\
    std. error & 10.06 & 7.54  &       & 2.15  & 5.03 \\
    \rowcolor[rgb]{ .875,  .89,  .898} t-stat & (4.556) & (3.116) &       & (7.328) & (2.661) \\
    rep\_score & 49.3*** & 23.5  &       & 20.9*** & 22.36*** \\
    \rowcolor[rgb]{ .875,  .89,  .898} std. error & 10.83 & 7.60  &       & 2.64  & 6.15 \\
    t-stat & (4.55) & (3.086) &       & (7.881) & (2.867) \\
    \midrule
    \rowcolor[rgb]{ .875,  .89,  .898} Wald test & 1.587 & 0.000391 &       & 24.239 & 3.315 \\
    p-value & 0.2077 & 0.9842 &       & 0.0000 & 0.0686 \\
    \midrule
    \rowcolor[rgb]{ .875,  .89,  .898} No. of obs. & 199650 & 59212 &       & 314376 & 49440 \\
    \bottomrule
    \end{tabular}%
  \caption{\textbf{Stacked regressions to compare the coefficients of the original GPT score and the replaced GPT score for statistically significant differences.} The formula of the regression is given by equation (\ref{eq: stacked1}).  Standard errors are clustered by time and firm. Each Wald test tests equality of the corresponding orig\_score and rep\_score coefficients.}
    \label{tab: regression2}
\end{table}%

\subsection{Comparing Market Betas of the Strategies}


Investors generally prefer strategies with lower correlation to overall market performance. Lower correlation offers greater diversification, and it suggests that a strategy's returns are driven by information unrelated to the overall market. We are therefore interested in comparing the performance of the original and replaced strategies through regression on the market. 
We run regressions of the form
%
\begin{equation}
    r_{t+1} = \beta_0 + \beta_1 (rm-rf)_t + \epsilon_{t+1},
    \label{capm}
\end{equation}
where $r_{t+1}$ is a portfolio return from $t$ to $t+1$, and $(rm-rf)_t$ is the market return minus the risk-free rate at time $t$, obtained from the Kenneth French data library \citet{ken}. We measure returns in basis points.

Given that our trading strategies rely on GPT recommendations from real-time news headlines, our portfolio returns should generally not be highly correlated with the return of the market. For instance, if the overall market is trending downwards, our model can still profit by buying good-news stocks and shorting bad-news stocks, as long as there is some short-term price underreaction to news about individual stocks.

Table~\ref{tab: regression3} reports results for eight regressions of the form in (\ref{capm}), corresponding to the news source (scraped or TR), the time period (in-sample or out-of-sample), and the headline type (original or replaced). The market betas (the coefficients for rm-rf) are all small. More importantly, in every combination of source and period, the replaced R$^2$ is lower than the original R$^2$, and the replaced market beta is lower than the original market beta. In fact, in all cases, the replaced market beta fails to be statistically significantly different from zero at the 5\% level. 
The replaced strategy is less correlated with the market and less sensitive to the direction of the market.

The results in Table~\ref{tab: regression3} may also shed light on the relative influences of look-ahead bias and the distraction effect. If GPT-3.5 were able to exploit look-ahead bias effectively, its in-sample profits should be highly idiosyncratic, driven by stocks with company-specific good or bad news. But for both news sources we see that the in-sample beta is larger than the out-of-sample, suggesting, once again, that the distraction effect outweighs GPT's ability to trade on look-ahead bias. The lowest betas of all are found for the out-of-sample performance of the replaced strategy. This is precisely the case in which we expect the distraction effect to be weakest and the sentiment signal in headlines about individual companies to be most idiosyncratic.

\begin{table}[htbp]
  \centering
    \begin{tabular}{lrrrrrrrrr}
    \toprule
          & \multicolumn{4}{c}{Scraped} &       & \multicolumn{4}{c}{Thomson Reuters} \\
    \midrule
          & \multicolumn{2}{c}{\small In-Sample} & \multicolumn{2}{c}{\small Out-of-Sample} &       & \multicolumn{2}{c}{\small In-Sample} & \multicolumn{2}{c}{\small Out-of-Sample} \\
          & \multicolumn{1}{l}{\small Original} & \multicolumn{1}{l}{\small Replaced} & \multicolumn{1}{l}{\small Original} & \multicolumn{1}{l}{\small Replaced} &       & \multicolumn{1}{l}{\small Original} & \multicolumn{1}{l}{\small Replaced} & \multicolumn{1}{l}{\small Original} & \multicolumn{1}{l}{\small Replaced} \\
    \midrule
    \rowcolor[rgb]{ .875,  .89,  .898} \small const & \small12.77*** & \small14.74*** & \small18.78*** & \small14.13*** &  & \small9.09***  & \small11.44*** & \small3.77  & \small3.54 \\
    \footnotesize std. error & \footnotesize 3.917 & \footnotesize 4.389 &\footnotesize6.623 & \footnotesize6.384 &  & \footnotesize1.783 & \footnotesize2.007 & \footnotesize4.117 & \footnotesize4.979 \\
    \rowcolor[rgb]{ .875,  .89,  .898} \footnotesize t-stat & \footnotesize(3.26)  & \footnotesize(3.358) & \footnotesize(2.837) & \footnotesize(2.213) &   & \footnotesize(5.102) & \footnotesize(5.701) & \footnotesize(0.917) & \footnotesize(0.712) \\
    \small rm-rf & \small0.429*** & \small0.348* & \small0.273***  & \small0.0850  &       & \small0.317*** & \small0.0677  & \small0.220*** & \small0.0221 \\
    \rowcolor[rgb]{ .875,  .89,  .898} \footnotesize std. error & \footnotesize0.0420 & \footnotesize0.215 & \footnotesize0.0532 & \footnotesize0.0694 &       & \footnotesize0.0192 & \footnotesize0.0488 & \footnotesize0.0331 & \footnotesize0.0677 \\
    \footnotesize t-stat & \footnotesize(10.217) & \footnotesize(1.620) & \footnotesize(5.131) & \footnotesize(1.226) &    & \footnotesize(16.571) & \footnotesize(1.389) & \footnotesize(6.654) & \footnotesize(0.326) \\
    \midrule
    \rowcolor[rgb]{ .875,  .89,  .898} \small No. of obs. & \small 1699  & \small 1699  & \small 314   & \small314   &   & \small1695  & \small1695  & \small314   & \small314 \\
    \small R$^2$     & \small 0.058 & \small 0.033 & \small 0.078 & \small 0.009 &   & \small 0.14  & \small0.005 & \small0.124 & \small0.001 \\
    \bottomrule
    \multicolumn{5}{l}{\cellcolor{white}\footnotesize* p $<$ 0.10, ** p $<$ 0.05, *** p $<$ 0.01} \\
    \end{tabular}%
    \caption{\textbf{Regression results of overall long-short portfolio returns on the excess return of the market.} }
    \label{tab: regression3}
\end{table}%

Table \ref{tab: diff test} confirms that the difference between in-sample and out-of-sample market betas reported in Table \ref{tab: regression3} is statistically significant for both news sources. This comparison treats returns in the two time periods as independent of each other.

\begin{table}[htbp]
  \centering
    \begin{tabular}{lccc}
    \toprule
          & Scraped &       & Thomson Reuters \\
    \midrule
    \rowcolor[rgb]{ .875,  .89,  .898} abs. diff. in coefficients & 0.1557** &       & 0.973*** \\
    std. error & 0.0678  &       & 0.0382 \\
    \rowcolor[rgb]{ .875,  .89,  .898} t-stat & 2.30  &       & 2.55 \\
    p-value & 0.011 &       & 0.005 \\
    \bottomrule
    \end{tabular}%
  \caption{\textbf{Independent samples t-test comparing the coefficient of the excess market return in-sample and the coefficient of the excess market return out-of-sample.}}
  \label{tab: diff test}%
\end{table}%


\section{Discussion}
\label{s:discussion}

The power of large language models lies in their ability to learn from massive training data. This feature enables LLMs to make short-term stock price predictions from news text without having been trained to do so. But the use of massive training data also limits opportunities to design and backtest strategies that leverage LLMs. Any backtest that includes an LLM's training window is tainted by the LLM's exposure to information about subsequent events.

Our analysis shows that an LLM's in-sample knowledge can usefully be thought of in two parts, 
which we refer to as look-ahead bias and a distraction effect.
Look-ahead bias overstates the performance of trading strategies based on an LLM's predictions, but the distraction effect can be positive or negative.

To measure and mitigate the bias in LLM in-sample predictions, we proposed an anonymization procedure that removes a company's name and other identifying information from the news text provided to the LLM. Surprisingly, anonymization \emph{improves} the in-sample performance of a long-short trading strategy driven by GPT-3.5 sentiment analysis of news headlines. We confirm this pattern in two different collections of news headlines. We conclude that the overall distraction effect must be negative and that it outweighs the positive gains from look-ahead bias. GPT-3.5 does not do a good job exploiting look-ahead bias, and the distraction effect adversely interferes with its ability to evaluate the sentiment in news text. This pattern is particularly evident for larger companies --- the companies we expect to be most represented in the GPT training data.

Out-of-sample performance is immune to look-ahead bias but not to the distraction effect. We find some evidence, although only borderline statistically significant, that anonymization is helpful in this setting as well. In other words, anonymization may be useful in improving GPT's out-of-sample sentiment analysis as well as in removing in-sample bias.

\pagebreak
\printbibliography 
\pagebreak
\appendix

\section{Long-Only and Short-Only Trading Performance}

Tables~\ref{tab:long} and~\ref{tab:short} reproduce the long-short analysis in Table~\ref{tab:sig tests} for the long-only and short-only portfolios. The results are similar across the three tables; in particular, the signs of the t-stats are the same in all three cases. Put differently, in six out of the eight comparisons in Tables~\ref{tab:long} and~\ref{tab:short}, the replaced headlines outperform the original headlines, although most of the differences are not statistically significant.

In out-of-sample performance using scraped headlines, the short-only strategy performs better using original than replaced headlines, and the difference is borderline significant with a p-value of 0.075. The advantage of the original headlines may be due in part to the smaller size of many companies covered by the scraped headlines. If a stock is hard to borrow (as often happens with smaller stocks), we expect to see a weaker market reaction to bad news, as there will be less short selling. An LLM may implicitly learn that certain companies experience weaker short-term responses to bad news. Knowing the identity of a company in a headline may therefore provide indirectly useful information in predicting the market reaction to a headline. However, we do not see such an effect in the in-sample period, where the distraction effect from other information may be greater.

%

\begin{table}[htbp]
  \centering
    \begin{tabular}{clccccc}
    \toprule
          &       & \multicolumn{2}{c}{Scraped} &       & \multicolumn{2}{c}{Thomson-Reuters} \\
          &       & Original & Replaced &       & Original & Replaced \\
    \midrule
    \multirow{5}[4]{*}{{In-Sample}} & \cellcolor[rgb]{ .875,  .89,  .898}No. of obs. & \cellcolor[rgb]{ .875,  .89,  .898}1699 & \cellcolor[rgb]{ .875,  .89,  .898}1699 & \cellcolor[rgb]{ .875,  .89,  .898} & \cellcolor[rgb]{ .875,  .89,  .898}1695 & \cellcolor[rgb]{ .875,  .89,  .898}1695 \\
          & Mean  & 24.09 & 29.69 &       & 9.54  & 11.28 \\
          & \cellcolor[rgb]{ .875,  .89,  .898}Std. dev. & \cellcolor[rgb]{ .875,  .89,  .898}234.24 & \cellcolor[rgb]{ .875,  .89,  .898}277.88 & \cellcolor[rgb]{ .875,  .89,  .898} & \cellcolor[rgb]{ .875,  .89,  .898}124.45 & \cellcolor[rgb]{ .875,  .89,  .898}136.86 \\
\cmidrule{2-7}          & t-stat & \multicolumn{2}{c}{-1.47} &       & \multicolumn{2}{c}{-1.14} \\
          & \cellcolor[rgb]{ .875,  .89,  .898}p-value & \multicolumn{2}{c}{\cellcolor[rgb]{ .875,  .89,  .898}0.142} & \cellcolor[rgb]{ .875,  .89,  .898} & \multicolumn{2}{c}{\cellcolor[rgb]{ .875,  .89,  .898}0.254} \\
    \midrule
    \multirow{5}[4]{*}{{Out-of-Sample}} & No. of obs. & 314   & 314   &       & 314   & 314 \\
          & \cellcolor[rgb]{ .875,  .89,  .898}Mean & \cellcolor[rgb]{ .875,  .89,  .898}9.78 & \cellcolor[rgb]{ .875,  .89,  .898}0.45 & \cellcolor[rgb]{ .875,  .89,  .898} & \cellcolor[rgb]{ .875,  .89,  .898}7.54 & \cellcolor[rgb]{ .875,  .89,  .898}14.32 \\
          & Std. dev. & 209.63 & 259.33 &       & 158.72 & 184.76 \\
\cmidrule{2-7}          & \cellcolor[rgb]{ .875,  .89,  .898}t-stat & \multicolumn{2}{c}{\cellcolor[rgb]{ .875,  .89,  .898}1.14} & \cellcolor[rgb]{ .875,  .89,  .898} & \multicolumn{2}{c}{\cellcolor[rgb]{ .875,  .89,  .898}-1.45} \\
          & p-value & \multicolumn{2}{c}{0.25} &       & \multicolumn{2}{c}{0.147} \\
    \bottomrule
    \end{tabular}%
  \caption{\textbf{Summary statistics and significance tests comparing long-only portfolios advised by original headlines and replaced headlines.} Each observation represents the average return of the portfolio on a given trading day.}
  \label{tab:long}
\end{table}%

\begin{table}[htbp]
  \centering
    \begin{tabular}{clccccc}
    \toprule
          &       & \multicolumn{2}{c}{Scraped} &       & \multicolumn{2}{c}{Thomson-Reuters} \\
          &       & Original & Replaced &       & Original & Replaced \\
    \midrule
    \multirow{5}[4]{*}{{In-Sample}} & \cellcolor[rgb]{ .875,  .89,  .898}No. of obs. & \cellcolor[rgb]{ .875,  .89,  .898}1699 & \cellcolor[rgb]{ .875,  .89,  .898}1699 & \cellcolor[rgb]{ .875,  .89,  .898} & \cellcolor[rgb]{ .875,  .89,  .898}1695 & \cellcolor[rgb]{ .875,  .89,  .898}1695 \\
          & Mean  & 30.43 & 37.36 &       & 14.96 & 18.25 \\
          & \cellcolor[rgb]{ .875,  .89,  .898}Std. dev. & \cellcolor[rgb]{ .875,  .89,  .898}265.69 & \cellcolor[rgb]{ .875,  .89,  .898}278.64 & \cellcolor[rgb]{ .875,  .89,  .898} & \cellcolor[rgb]{ .875,  .89,  .898}163.18 & \cellcolor[rgb]{ .875,  .89,  .898}161.97 \\
\cmidrule{2-7}          & t-stat & \multicolumn{2}{c}{-1.49} &       & \multicolumn{2}{c}{-1.82} \\
          & \cellcolor[rgb]{ .875,  .89,  .898}p-value & \multicolumn{2}{c}{\cellcolor[rgb]{ .875,  .89,  .898}0.137} & \cellcolor[rgb]{ .875,  .89,  .898} & \multicolumn{2}{c}{\cellcolor[rgb]{ .875,  .89,  .898}0.069} \\
    \midrule
    \multirow{5}[4]{*}{{Out-of-Sample}} & No. of obs. & 314   & 314   &       & 314   & 314 \\
          & \cellcolor[rgb]{ .875,  .89,  .898}Mean & \cellcolor[rgb]{ .875,  .89,  .898}22.99 & \cellcolor[rgb]{ .875,  .89,  .898}10.02 & \cellcolor[rgb]{ .875,  .89,  .898} & \cellcolor[rgb]{ .875,  .89,  .898}3.52 & \cellcolor[rgb]{ .875,  .89,  .898}12.17 \\
          & Std. dev. & 248.54 & 257.23 &       & 208.81 & 201.00 \\
\cmidrule{2-7}          & \cellcolor[rgb]{ .875,  .89,  .898}t-stat & \multicolumn{2}{c}{\cellcolor[rgb]{ .875,  .89,  .898}1.79} & \cellcolor[rgb]{ .875,  .89,  .898} & \multicolumn{2}{c}{\cellcolor[rgb]{ .875,  .89,  .898}-2.15} \\
          & p-value & \multicolumn{2}{c}{0.075} &       & \multicolumn{2}{c}{0.032} \\
    \bottomrule
    \end{tabular}%
  \caption{\textbf{Summary statistics and significance tests comparing short-only portfolios advised by original headlines and replaced headlines.} Each observation represents the average return of the portfolio on a given trading day.}
  \label{tab:short}
\end{table}%

\pagebreak

\end{document}